# Bose Hubbard Models with Synthetic Spin-Orbit Coupling: Mott Insulators, Spin Textures and Superfluidity

William S. Cole,[1] Shizhong Zhang,[1] Arun Paramekanti,[2,3] and Nandini Trivedi[1]

[1] *Department of Physics, The Ohio State University, Columbus OH 43210, USA*
[2] *Department of Physics, University of Toronto, Toronto M5S1A7, Canada*
[3] *Canadian Institute for Advanced Research, Toronto, Ontario, M5G 1Z8, Canada*
(Dated: May 10, 2012)

Motivated by the experimental realization of synthetic spin-orbit coupling for ultracold atoms, we investigate the phase diagram of the Bose Hubbard model in a non-abelian gauge field in two dimensions. Using a strong coupling expansion in the combined presence of spin-orbit coupling and tunable interactions, we find a variety of interesting magnetic Hamiltonians in the Mott insulator (MI), which support magnetic textures such as spin spirals and vortex and Skyrmion crystals. An inhomogeneous mean field treatment shows that the superfluid (SF) phases inherit these exotic magnetic orders from the MI and display, in addition, unusual modulated current patterns. We present a slave boson theory which gives insight into such intertwined spin-charge orders in the SF, and discuss signatures of these orders in Bragg scattering, *in situ* microscopy, and dynamic quench experiments.

*Introduction.*—Strong spin-orbit (SO) interaction is the key to realizing such remarkable states of electronic matter as topological band insulators [1, 2] and Weyl semimetals [3]. SO coupled Mott insulators can also realize the Kitaev model [4] which may enable the study of Majorana fermions in a condensed matter setting and provide a platform for topological quantum computation [5]. This has motivated parallel experimental advances in ultracold atomic gases, where Raman processes can be used to create tunable SO coupling, or more general nonabelian gauge fields [6–8], thus paving the way for investigating SO coupling and its emergent consequences for atomic fermions as well as bosons.

Experiments [6–9] and theory [10–15] on such SO coupled bosons have, so far, mainly focused on Bose-Einstein condensation in *weakly* interacting gases in the absence of a lattice. However, as theory [16–18] and experiments [19] in the absence of SO interaction have shown, tuning the lattice depth for bosons in an optical lattice can lead to a strongly interacting regime, accompanied by a suppression of the condensate density and finally a quantum phase transition into a featureless Mott insulator [20]. By contrast, the physics of SO coupled atoms in a *strongly* interacting regime and in an *optical lattice*, both of which are expected to lead to unique phenomena, remains a relatively unexplored frontier [21].

One of the most significant results in this Letter is our discovery that tuning SO coupling and interparticle interactions for 'spinful' bosons at a filling of one boson per site, leads to Mott insulating states with a *plethora* of magnetic exchange Hamiltonians including Dzyaloshinskii-Moriya (DM) interactions [22, 23]. This provides a toolbox to simulate a wide class of interesting quantum magnetic Hamiltonians including quantum compass models. These magnetic Hamiltonians on a two-dimensional (2D) square lattice are shown to have a rich classical phase diagram, exhibiting Ising and XY ordered ferromagnets, an Ising antiferromagnetic phase, two types of spiral phases, and vortex and Skyrmion crystals. We note here that compared with solid state materials, it is much easier to tune across this phase diagram by varying the relative importance of the DM interaction with respect to the exchange interaction. Upon increasing the boson tunneling, we find emergent superfluid phases that inherit magnetic textures from the Mott insulator state. We formulate a slave boson approach that provides a unified understanding of such intertwined spin-charge orders in the SF phase, as well as a description of the SF-MI transitions. We conclude with a discussion of specific experimental predictions which emerge from our theory.

*Model.*—We consider bosons with two hyperfine states ($\uparrow$ and $\downarrow$), described by the following Hamiltonian on a 2D square lattice:

$$H = -t\sum_{\langle ij\rangle}(\psi_i^\dagger \mathcal{R}_{ij}\psi_j + \text{h.c.}) + \frac{1}{2}\sum_{i\sigma\sigma'} U_{\sigma\sigma'} a_{i\sigma}^\dagger a_{i\sigma'}^\dagger a_{i\sigma'} a_{i\sigma} \quad (1)$$

where $\psi_i^\dagger = (a_{i\uparrow}^\dagger, a_{i\downarrow}^\dagger)$, and $a_{i\sigma}^\dagger$ creates a spin-$\sigma$ boson at site $i$. The first term above describes tunneling of bosons between neighboring sites, with $t$ being the overall hopping amplitude. The matrix $\mathcal{R}_{ij} \equiv \exp[i\vec{A}\cdot(\vec{r}_i - \vec{r}_j)]$, where $\vec{A} = (\alpha\sigma_y, \beta\sigma_x, 0)$ is a static, *non-abelian*, background gauge field seen by the lattice bosons. Diagonal terms in this matrix describe the usual spin-conserving hopping of bosons, while off-diagonal spin-flip terms describe the SO coupling which arises from a suitable two-photon Raman process [24]. We set $\beta = -\alpha$, for which the SO coupling is the lattice analog of the well-known Rashba term. The second term describes boson interactions; we choose the intraspecies repulsion $U_{\uparrow\uparrow} = U_{\downarrow\downarrow} \equiv U$, and set the interspecies interaction $U_{\uparrow\downarrow} = U_{\downarrow\uparrow} \equiv \lambda U$.

We analyze this model using various methods: (i) a weak coupling ($U, \lambda U \ll t$) Gross-Pitaevskii approach to study the condensate structure, (ii) a strong coupling $(U,\lambda U\gg t)$ approach to understand the Mott state and associated spin textures, (iii) an inhomogeneous mean field theory to describe the emergent strongly correlated superfluids, and (iv) a slave boson theory to gain insights into the coupled magnetic and charge orders.

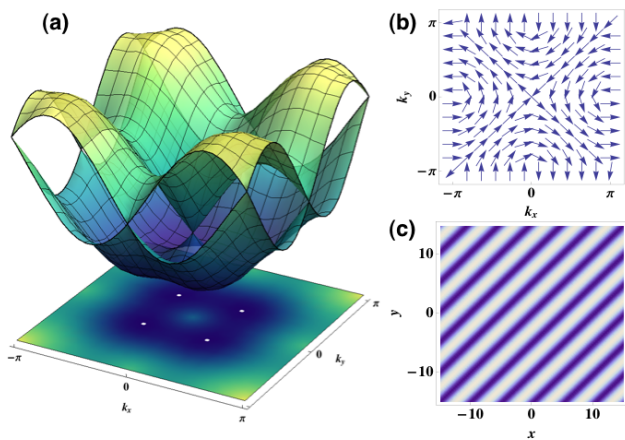

FIG. 1. (Color online) (a) The band structure of the hamiltonian in Eq. (1) for $U=0$, with $\alpha=-\beta=\pi/4$. There are four degenerate points at $\vec{Q}_1,\cdots\vec{Q}_4$ in the lower Bloch band due to the rotational symmetry breaking by the underlying square optical lattice, with $\tan k_0 = (\tan\alpha)/\sqrt{2}$, and a Dirac cone at the $\Gamma$ point. (b) The spin orientations in the lower Bloch band. The spin is locked to the momentum through the SO coupling. (c) Real-space stripe density distribution of spin-up particles in the condensate with $\lambda=1.3$ from the mean field GP calculation. There is a similar distribution for the spin-down particles, but with a $\pi/k_0$ shift perpendicular to the stripe direction. On the other hand, for $\lambda<1$, the spin density is uniform. The *total* density is uniform for all $\lambda$.

TABLE I. Exchange couplings in the effective hamiltonian. By taking $\alpha$ and $\lambda$ as tunable parameters a plethora of quantum magnetic Hamiltonians can be realized.

| | |
|---|---|
| $J_{\hat{x}}^{x} = -\frac{4t^2}{\lambda U}\cos(2\alpha)$ | $J_{\hat{y}}^{x} = -\frac{4t^2}{\lambda U}$ |
| $J_{\hat{x}}^{y} = -\frac{4t^2}{\lambda U}$ | $J_{\hat{y}}^{y} = -\frac{4t^2}{\lambda U}\cos(2\alpha)$ |
| $J_{\hat{x}}^{z} = -\frac{4t^2}{\lambda U}(2\lambda-1)\cos(2\alpha)$ | $J_{\hat{y}}^{z} = -\frac{4t^2}{\lambda U}(2\lambda-1)\cos(2\alpha)$ |
| $\vec{D}_{\hat{x}} = -\frac{4t^2}{U}\sin(2\alpha)\hat{y}$ | $\vec{D}_{\hat{y}} = \frac{4t^2}{U}\sin(2\alpha)\hat{x}$ |

*Weak coupling superfluid.*—The non-interacting band structure for the above model shown in Fig. 1(a) has four degenerate minima in the lowest Bloch band at $\vec{Q}_1 = (k_0,k_0)$, $\vec{Q}_2 = (-k_0,k_0)$, $\vec{Q}_3 = (-k_0,-k_0)$ and $\vec{Q}_4 = (k_0,-k_0)$, where $\tan k_0 = (\tan\alpha)/\sqrt{2}$. This is in stark contrast to the continuum case where the minima form a degenerate circle, and suggests that Rashba coupled Bose condensates confined to an optical lattice are more stable against fluctuations than their continuum counterparts. We label the Bloch eigenstates at these points as $\varphi_m = \exp(i\vec{Q}_m\cdot\vec{r})\chi_m$, $m=1,\ldots,4$. The spin wavefunction $\chi_m$ associated with $\varphi_m$ has the form $\chi_m^\dagger \equiv (1/\sqrt{2})(1,\exp(-im\pi/4))$. More generally, the spin wavefunction winds around the $\Gamma$ point in the first Brillouin zone with a winding number 1, as shown in Fig. 1(b).

Within the Gross-Pitaevskii (GP) approximation, all $N$ bosons condense into a common single particle state $\Phi = \sum_m c_m \varphi_m$ where $c_m$ are complex variational parameters, satisfying $\sum_m |c_m|^2=1, m=1,\cdots 4$. Setting $\Phi^\dagger \equiv (\Phi_\uparrow^*,\Phi_\downarrow^*)$, we determine $c_m$ by minimizing the interaction energy $U_{\rm int}(\{c_m\}) \equiv NU/2(|\Phi_\uparrow|^4+|\Phi_\downarrow|^4+2\lambda|\Phi_\uparrow|^2|\Phi_\downarrow|^2)$. As an illustration, for $\alpha = \pi/4$ we find the following structure of the condensate: For $\lambda<1$, only one of the lowest four degenerate states is occupied. In this case, both the spin and number density of the superfluid are uniform, and the ground state is four-fold degenerate. On the other hand, for $\lambda>1$, two of the lowest four states with opposite wave vectors are occupied. This leads to stripe order in the spin density (see Fig. 1(c)) while the total density remains uniform. The wave vector corresponding to the spin-stripe density is $2\sqrt{2}k_0$ and the ground state is two-fold degenerate. As we will see below, such magnetic states are also found at strong coupling, where however this GP approach focusing on just the four minima at $\vec{Q}_m$ misses additional magnetic textures.

*Strong coupling Mott phases.*—At unit filling and for $U/t=\infty$, the repulsive boson interactions favor exactly one boson at each site. The ground states at $t=0$ are highly degenerate, with an arbitrary spin state at each site. Away from this limit, to $\mathcal{O}(t^2/U)$, we obtain the effective low-energy spin Hamiltonian

$$H_{\rm spin} = \sum_{i,\delta=\hat{x},\hat{y}} \left\{ \sum_{a=x,y,z} J_\delta^a S_i^a S_{i+\delta}^a + \vec{D}_\delta \cdot (\vec{S}_i \times \vec{S}_{i+\delta}) \right\} \quad (2)$$

where the exchange coupling constants $J_\delta^a$ and DM vectors $\vec{D}_\delta$ are given in Table I. Thus, simply by tuning $\alpha$ and $\lambda$ in a single system, one can emulate a plethora of Hamiltonians of great interest in quantum magnetism. As an example, for $\alpha=0$, $H_{\rm spin}$ reduces to an XXZ magnet [25] with negative (ferromagnetic) $xy$-coupling and a $z$-coupling determined by $(1-2\lambda)$. For $\alpha\neq 0$, one obtains both anisotropic exchange couplings as well as a DM interaction which tends to induce spin spirals as in chiral magnets like MnSi. For $\alpha=\pi/4$ we find a "compass"-type model with a DM perturbation. The Hamiltonian in Eq. (1) thus constitutes perhaps the simplest itinerant model with demonstrably chiral magnetic ground states.

We obtain the classical ground state phase diagram of $H_{\rm spin}$ in Eq.(2) via Monte Carlo annealing methods [26] (see Fig. 2). We find the following phases, and characterize them by their magnetic structure factors $S_{\vec{q}} = |\sum_i \vec{S}_i e^{i\vec{q}\cdot\vec{r}_i}|$.

<u>xyFM/zFM</u>: In these ferromagnetic phases, the spin structure factor exhibits a peak at $\vec{q}=(0,0)$. The zFM has spins along the $\pm z$-axis. In the xyFM, the SO interaction pins the spins to lie in the $xy$-plane making angles $(2n+1)\pi/4$ (with $n=0\ldots3$) with the $x$-axis.

<u>zAFM</u>: In the zAFM, $S_{\vec{q}}$ exhibits a peak at $(\pi,\pi)$, with spins pointing along the $\pm z$-axis.

<u>Spiral-1</u>: This is a coplanar ground state, with the spins spiralling in the plane defined by the vectors $\hat{z}$-$\vec{q}$, where $\vec{q}\equiv(q,\pm q)$ is an incommensurate wavevector.

<u>Spiral-2</u>: This is a coplanar ground state, with the spins spiralling in the $\hat{z}$-$\vec{q}$ plane, where $\vec{q}\equiv(q,0)$ (or $(0,q)$) is incommensurate for small $\alpha$, but there is a region of the phase diagram (light green region of "Spiral-2" in Fig. 2) that supports a commensurate $(4\times1)$-site unit cell.

<u>$2\times2$ Vortex Crystal (VX)</u>: This is a coplanar ground state, with spins in the $xy$-plane having components $S_x=(-1)^x/\sqrt{2}$ and $S_y=(-1)^y/\sqrt{2}$. Thus the spins wind clockwise or counterclockwise around each plaquette. The VX has $S_{\vec{q}}$ peaks at $(\pi,0)$ and $(0,\pi)$.

<u>$3\times3$ Skyrmion Crystal (SkX)</u>: This is a <i>non</i>-coplanar state, where the spins form a $3\times3$ unit cell that has nonzero Skyrmion density, given by $\sum_i \vec{S}_i\cdot(\vec{S}_{i+\hat{x}}\times\vec{S}_{i+\hat{y}})$. The structure factor has peaks at $(2\pi/3,0)$ and $(0,2\pi/3)$.

The Spiral-1, Spiral-2, VX, and SkX phases break the $C_{4v}$ symmetry of the square lattice; they are thus expected to undergo multiple thermal transitions, associated with restoring spin rotational and lattice rotational symmetries, enroute to the high temperature paramagnetic state. In the Spiral-1 phase, for example, these multiple transitions are manifested through two specific heat peaks in our classical Monte Carlo simulations [27].

<i>Mott lobes and magnetically textured SFs.</i>—In order to address the strongly correlated superfluid phase beyond the GP approach, and to connect with the magnetic textures in the Mott insulator, we extend the usual "single site" mean field theory [28] to spinful bosons. In order to capture the non-uniform magnetic ordering, and possible inhomogeneous superfluidity, we decompose the kinetic term of the Hamiltonian in Eq.(1) to allow for a spatially varying condensate order parameter $\phi_{i\sigma}=\langle a_{i\sigma}\rangle$. The self-consistent solution of this mean field theory requires an iterative minimization over a finite cluster [26].

For $t=0$, the single site Hamiltonian is $H_U=(U/2)(n_\uparrow^2+n_\downarrow^2+2\lambda n_\uparrow n_\downarrow)-(\mu+U/2)(n_\uparrow+n_\downarrow)$. We determine the maximum size of the Mott lobe to be $\min(U,\lambda U)$ along the $\mu$-axis. As we increase $t/U$, there is a quantum phase transition from the magnetic insulating states to the superfluid states at a critical value $(t/U)_c$, which increases with $\alpha$ for fixed $\lambda$. This is consistent with previous results obtained using a hopping expansion [21] which, however, only addressed the homogeneous Mott phase with xyFM magnetic order.

To characterize the magnetic structures in the superfluid phase, we calculate (i) the local magnetic mo-ment $\vec{m}_i\equiv\langle a_{i\mu}^\dagger\vec{\sigma}_{\mu\nu}a_{i\nu}\rangle$ and (ii) the bond current $\kappa_{ij}^{\mu\nu}=-it(\mathcal{R}_{ij}^{\mu\nu}\langle a_{i\mu}^\dagger a_{j\nu}\rangle-\text{c.c.})$, where $ij$ are nearest neighbors. For the phases we find, the diagonal term $\mu=\nu$ of $\kappa_{ij}^{\mu\nu}$ is zero, while the nonzero off-diagonal term represents the total current arising from spin flip processes.

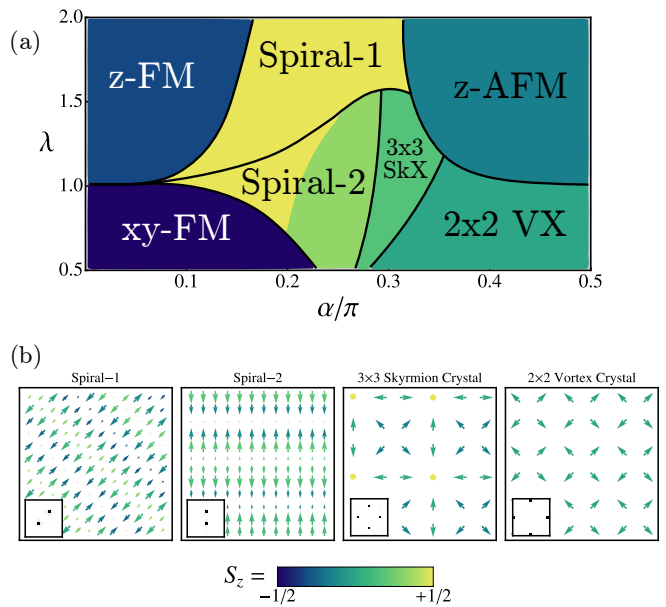

FIG. 2. (Color online) (a) Novel magnetic phases in the Mott insulating regime, obtained from Monte Carlo annealing of the spin hamiltonian Eq. (2). Spin configurations are abbreviated as follows. zFM (xyFM) denotes a ferromagnet with magnetic moment along $\hat{z}$-direction (in the $xy$-plane). zAFM denotes an anti-ferromagnet with staggered moments along the $\hat{z}$-direction. The Spiral-1 phase has spiral vector along the $(\pi,\pi)$ direction and Spiral-2 has a spiral vector along the $(\pi,0)$ direction. The green area of the Spiral-2 region represents a commensurate 4-site spiral. VX and SkX denote a vortex crystal and a Skyrmion crystal phase with four Bragg spots. (b) shows the $xy$-plane projection of the real space spin configurations in the Spiral-1, 2, SkX, and VX phases. The magnetic structure factor peaks (as described above) are shown in the insets.

In Fig.3 (A) and (B), we plot the Mott lobes for filling $n=1$ and $\alpha=\pi/2$, together with the $z$-component of the site spin density and bond currents in the SF phase. We find, as shown, that for $\lambda=1.5$ and $\lambda=0.5$ the magnetic order in the SF reflects the magnetic ordering in the Mott state from which they emerge. In addition, however, the SFs support plaquette currents which form a checkerboard pattern. We find that for $\lambda=1.5$, where the strong coupling phase is a zAFM, this current order spontaneously breaks the time reversal symmetry (in picking one of the two allowed checkerboard patterns), while for $\lambda=0.5$, the underlying magnetic phase picks a unique loop current order. To understand this interplay between magnetic order and unusual bond current patterns found in our inhomogeneous mean field theory, we next formulate a slave boson theory of this problem which also provides a unified framework to understand the SF and MI phases and the SF-MI transitions.

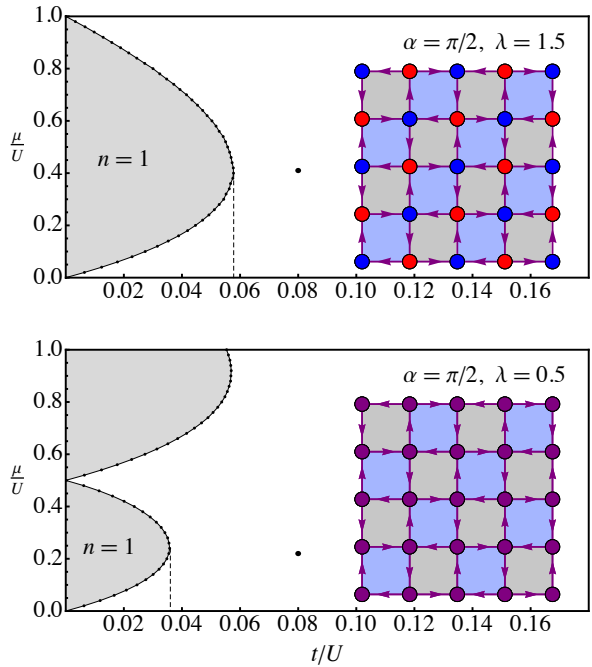

FIG. 3. (Color online) Phase diagrams of the spin-orbit coupled Bose-Hubbard model in $\mu/U$ vs. $t/U$ plane, showing Mott lobes and superfluid states. (A) phase diagram with $\lambda = 1.5$ and $\alpha = \pi/2$ and (B) $\lambda = 0.5$ and $\alpha = \pi/2$. The width of the $n = 1$ lobe is given by $\lambda U$ and the critical value $(t/U)_c$ increases with $\lambda$. The two insets show the local spin density distribution (red=↑, blue=↓, purple in between) and bond currents for $t/U = 0.08$ in the superfluid phase close to the Mott states. $\mu/U$ is tuned such that the average number of particles per site is unity. For $\lambda = 1.5$, the spin density assumes $z$-antiferromagnetic order whereas for $\lambda = 0.5$, the magnetic moments are in the $2 \times 2$ VX phase, and restricted to the $xy$-plane. The bond currents for both $\lambda = 1.5$ and $\lambda = 0.5$ share the same pattern, with clockwise and anti-clockwise plaquette loop currents forming Ising anti-ferromagnetic order.

*Slave boson theory.*—Inspired by theories of strongly correlated electronic materials [29, 30], we express the Hubbard model in terms of separate bosonic spin and charge degrees of freedom by setting $a_{i\sigma} = f_{i\sigma} b_i$, where the $b$-bosons (chargons) carry charge but no spin, while the $f$-bosons (spinons) carry spin but no charge. To remain in the physical Hilbert space, we impose the local constraint $b_i^\dagger b_i = \sum_\sigma f_{i\sigma}^\dagger f_{i\sigma}$. At mean-field level, where this constraint is treated on average, we are led to two separate but coupled Hamiltonians for the spinons and chargons which need to be solved self-consistently [26].

Here, our goal is to understand the superfluid phases with magnetic textures derived from the Mott phase as indicated by the mean field analysis given above. Such magnetic textures can be obtained by condensing the spinons into an appropriate condensate wavefunction $\Phi_{i\sigma}$. This leads to an effective chargon Hamiltonian

$$H_b = -t \sum_{\langle ij \rangle \mu\nu} (R_{ij}^{\mu\nu} \Phi_{i\mu}^* \Phi_{j\nu} b_i^\dagger b_j + \text{h.c.}) + \frac{U}{2} \sum_i b_i^\dagger b_i^\dagger b_i b_i. \quad (3)$$

Different magnetic textures thus 'act' on the chargons as distinct effective *abelian* gauge field configurations. We show below that this unusual charge-spin coupling is the origin of the numerically observed bond current patterns in the SF as shown in Fig.3.

To begin with a simple example, consider the magnetic order in the zFM which is captured by setting $\Phi_{i\uparrow} = 1$ and $\Phi_{i\downarrow} = 0$. This leads to a conventional Bose Hubbard model for the chargons, but with a renormalized chargon hopping amplitude $t \cos\alpha$. This renormalizes the critical interaction needed to drive the Mott transition at unit filling from $U_c^0$ for spinless bosons to $U_c^{zFM}(\alpha) = U_c^0 \cos\alpha$. Going beyond mean field theory, we conclude that this Mott transition remains in the 3D-XY universality class [27].

To understand the bond current ordered SF emerging from the insulator with zAFM order, we set $\Phi_{\mathbf{r}\uparrow} = 1$ and $\Phi_{\mathbf{r}\downarrow} = 0$ on the A sublattice, and $\Phi_{\mathbf{r}\downarrow} = 1$ and $\Phi_{\mathbf{r}\uparrow} = 0$ on the B sublattice. In contrast to the previous case, the chargon Hamiltonian is now found to enclose $\pi$-flux per plaquette for the chargons. This flux results in the spontaneous checkerboard pattern [31, 32] of mass currents seen in Fig.3 (A). The SF phases emerging from the SkX, VX, Spiral-1 and Spiral-2 phases, as well as a complete SB mean field theory, treating magnetic and charge orders self-consistently, will be discussed elsewhere [27].

*Experimental implications.*—One of the most interesting aspects of our work is the realization that one can tune across a wide variety of magnetic Hamiltonians, which support magnetically textured Mott insulators and superfluids, starting from the simple Bose Hubbard model in Eq. (1). The magnetic structure factor $S_{\vec{k}}$ in the different phases, shown in Fig.2, can be directly measured with optical Bragg scattering experiments [33].

An alternative route to exploring the magnetic MI and SF phases is via *in situ* microscopy which has the demonstrated ability to detect lattice-resolved hyperfine states and number fluctuations of atoms [34]. Finally, the unusual bond currents in the SF phases, such as the checkerboard current pattern in the SF phase descending from the zAFM, could be detected using a recent proposal to quench the lattice potential along one direction which dynamically converts such atomic current patterns into measurable atomic density patterns [35]. Such experiments would lead us to a deeper understanding of the emergent consequences of the interplay of spin-orbit coupling and strong interactions for bosons.

We thank J. Radić, A. Di Ciolo, K. Sun, and V. Galitski for discussions and comparing data on the classical magnetic phases. We acknowledge funding from ARO Grant W911NF-08-1-0338 (W.S.C), DARPA under the Optical Lattice Emulator program R15835 (S.Z.) and NSF DMR-0907275 (N.T.). A.P. acknowledges support from NSERC (Canada).magnetic phases. We acknowledge funding from ARO Grant W911NF-08-1-0338 (W.S.C), DARPA under the Optical Lattice Emulator program R15835 (S.Z.) and NSF DMR-0907275 (N.T.). A.P. acknowledges support from NSERC (Canada).

# Supplementary Material for "Bose Hubbard Model with Synthetic Spin Orbit Coupling: Mott Insulators, Spin Textures and Superfluidity"

William S. Cole[1], Shizhong Zhang[1], Arun Paramekanti[2,3], Nandini Trivedi[1]
[1]*Department of Physics, The Ohio State University, Columbus OH 43210, USA*
[2]*Department of Physics, University of Toronto, Toronto M5S1A7, Canada and*
[3]*Canadian Institute for Advanced Research, Toronto, Ontario, M5G 1Z8, Canada*

## SLAVE BOSON THEORY OF BOSE HUBBARD MODEL WITH SPIN-ORBIT COUPLING

**Hamiltonian:** Consider two-component bosons with a Hamiltonian the form

$$H = -t \sum_{\mathbf{r},\delta=\hat{x},\hat{y}} (a^\dagger_{\mathbf{r},\mu} R^{\mu\nu}_\delta a_{\mathbf{r}+\delta,\nu} + \text{h.c.}) - \mu \sum_{\mathbf{r},\sigma} n_{\mathbf{r}\sigma} + \frac{U}{2} \sum_{\mathbf{r}} (n^2_{\mathbf{r}\uparrow} + n^2_{\mathbf{r}\downarrow} + 2\lambda n_{\mathbf{r}\uparrow} n_{\mathbf{r}\downarrow}) \quad (1)$$

where $R_x = \cos\alpha - i\sin\alpha\sigma^y$ and $R_y = \cos\alpha + i\sin\alpha\sigma^x$, and we assume the chemical potential is tuned to be at a filling of one boson per site on average (counting both spins).

**Slave boson formulation:** To capture the Mott transition of such spin-orbit coupled superfluids, it is useful to formulate a slave boson mean field theory by splitting the spin and charge degrees of freedom of the bosons. If we assume that charges can Bose condense to yield a superfluid or localize into Mott phase, while the spins order both in the Mott and superfluid phases, such a formulation is capable of capturing both magnetically ordered superfluids and insulators found via extended mean field theory or strong coupling expansions respectively. It can also be suitably generalized to study magnetically disordered 'spin-liquid' Mott insulators. To be concrete, let us consider a slave particle decomposition of the original spinful boson, as

$$a_{\mathbf{r},\sigma} = b_\mathbf{r} f_{\mathbf{r},\sigma} \quad (2)$$

where $b$ (chargon) and $f$ (spinon) are both bosons which obey the local constraint $b^\dagger_\mathbf{r} b_\mathbf{r} = \sum_\sigma f^\dagger_{\mathbf{r},\sigma} f_{\mathbf{r},\sigma}$. Here the chargons carry charge but no spin, while the spinons carry spin but no charge. (The "charge" here refers to the number of bosons.) In addition, we see that we can rotate $b_\mathbf{r} \to b_\mathbf{r} e^{i\chi_\mathbf{r}}$ and $f_{\mathbf{r}\sigma} \to f_{\mathbf{r}\sigma} e^{-i\chi_\mathbf{r}}$ which leaves the original boson operator $a_{\mathbf{r}\sigma}$ invariant. This means that the spinons and chargons carry opposite U(1) "gauge charge", and are coupled by a U(1) gauge field which serves to impose the local constraint, as is standard in such slave particle theories. To begin, we will consider only classical magnetically ordered states in this study, for which it suffices to set $f_{\mathbf{r}\sigma} \to \Phi_{\mathbf{r}\sigma}$, where $\Phi_{\mathbf{r}\sigma}$ is a complex number; this amounts to completely condensing the spinons. Since the spinons are condensed, the dynamical U(1) gauge field is gapped out by the Higgs mechanism, and we expect a mean field description of the resulting phases to be a reliable starting point.

**Slave boson mean field theory:** In slave boson language, the Hamiltonian takes the form

$$\begin{aligned}H = & -t \sum_{\mathbf{r},\delta=\hat{x},\hat{y}} (b^\dagger_\mathbf{r} b_{\mathbf{r}+\delta} f^\dagger_{\mathbf{r},\mu} R^{\mu\nu}_\delta f_{\mathbf{r}+\delta,\nu} + \text{h.c.}) - \mu \sum_{\mathbf{r},\sigma} f^\dagger_{\mathbf{r}\sigma} f_{\mathbf{r}\sigma} - \sum_\mathbf{r} \mu^c_\mathbf{r}(b^\dagger_\mathbf{r} b_\mathbf{r} - \sum_\sigma f^\dagger_{\mathbf{r}\sigma} f_{\mathbf{r}\sigma}) \\ & + \frac{U}{2} \sum_\mathbf{r}(n^2_{f,\mathbf{r}\uparrow} + n^2_{f,\mathbf{r}\downarrow} + 2\lambda n_{f,\mathbf{r}\uparrow} n_{f,\mathbf{r}\downarrow}).\end{aligned} \quad (3)$$

where $\mu$ tunes the overall density, such that $\sum_\sigma \langle n_{f,\mathbf{r}\sigma}\rangle = \sum_\sigma \langle a^\dagger_{\mathbf{r}\sigma} a_{\mathbf{r}\sigma}\rangle$, while $\mu^c_\mathbf{r}$ enforces the local *average* constraint $\langle b^\dagger_\mathbf{r} b_\mathbf{r}\rangle = \sum_\sigma \langle f^\dagger_{\mathbf{r},\sigma} f_{\mathbf{r},\sigma}\rangle$. For now, let us limit ourselves to using this slave particle description to gain some insights into the various magnetically ordered superfluids and Mott insulators found from the mean field and strong coupling analyses. Minimizing the Hamiltonian, while assuming a simple 'classical' magnetic order, then amounts to finding a self-consistent ground state of the following two Hamiltonians:

$$H_b = -t \sum_{\mathbf{r},\delta=\hat{x},\hat{y}} (R^{\mu\nu}_\delta \Phi^*_{\mathbf{r},\mu} \Phi_{\mathbf{r}+\delta,\nu} b^\dagger_\mathbf{r} b_{\mathbf{r}+\delta} + \text{h.c.}) + \frac{U}{2}\sum_\mathbf{r} b^\dagger_\mathbf{r} b^\dagger_\mathbf{r} b_\mathbf{r} b_\mathbf{r} \quad (4)$$

$$H_f = -t \sum_{\mathbf{r},\delta=\hat{x},\hat{y}} (R^{\mu\nu}_\delta B_{\mathbf{r},\mathbf{r}+\delta} \Phi^*_{\mathbf{r},\mu} \Phi_{\mathbf{r}+\delta,\nu} + \text{c.c.}) + U(\lambda-1) \sum_\mathbf{r} |\Phi_{\mathbf{r}\uparrow}|^2 |\Phi_{\mathbf{r},\downarrow}|^2, \quad (5)$$

subject to the constraints $\frac{1}{N}\sum_{\mathbf{r}}\langle b_{\mathbf{r}}^\dagger b_{\mathbf{r}}\rangle = \bar{n}$ (where $\bar{n}=1$ is the average $a$-boson filling summed over both spins), while $\sum_\sigma |\Phi_{\mathbf{r},\sigma}|^2 = \langle b_{\mathbf{r}}^\dagger b_{\mathbf{r}}\rangle$ at each site $\mathbf{r}$. Here we have defined $B_{\mathbf{r},\mathbf{r}+\delta} = \langle b_{\mathbf{r}}^\dagger b_{\mathbf{r}+\delta}\rangle$, and $H_f$ is a purely classical energy functional. We have simplified the interaction part, writing

$$\frac{U}{2}(n_{f,\mathbf{r}\uparrow}^2 + n_{f,\mathbf{r}\downarrow}^2 + 2\lambda n_{f,\mathbf{r}\uparrow}n_{f,\mathbf{r}\downarrow}) = \frac{U}{2}(n_{f,\mathbf{r}\uparrow}+n_{f,\mathbf{r}\downarrow})^2 + U(\lambda-1)n_{f,\mathbf{r}\uparrow}n_{f,\mathbf{r}\downarrow} \tag{6}$$

$$= \frac{U}{2}n_{b,\mathbf{r}}^2 + U(\lambda-1)n_{f,\mathbf{r}\uparrow}n_{f,\mathbf{r}\downarrow}, \tag{7}$$

where we have used the exact constraint in the second line in rewriting it into the final form. Since we are imposing constraints only on average while solving this model, our choice for writing the interaction part is made to ensure that when we set $\alpha=0$ and assume a single species of bosons by setting $\Phi_{\mathbf{r}\sigma} = \delta_{\sigma,\uparrow}$ (say), the chargon Hamiltonian can reproduce the usual Mott transition of spinless bosons. To analyze the magnetically ordered studied using MC and extended mean field theory, and to understand its impact on superfluidity and the Mott transitions, we derive the chargon models obtained by fixing the magnetic order by condensing the spinons.

### DETAILS OF THE MONTE CARLO ANNEALING METHOD FOR CLASSICAL SPIN GROUND STATES

We treat the exchange model in a classical approximation where the spin operators are treated as classical unit vectors $\vec{S}_i = S(\sin\theta_i\cos\phi_i, \sin\theta_i\sin\phi_i, \cos\theta_i)$. The spins are first initialized to a random configuration, then random local updates are proposed on $(\theta_i, \phi_i)$ which are accepted according to the relative Boltzmann weight of the two configurations. We do simulated annealing with 50 independent initial configurations, taking $\beta J$ from 1 to 20 on each run. We perform $5\cdot 10^5$ lattice sweeps, consisting of one proposed local update per site, at each temperature. Therefore we have 50 different annealed configurations at each value of $\alpha$, which indicates a statistical spread, but we don't perform any averaging with respect to spin configurations, rather we just identify the lowest energy configurations.

After performing the previous steps, we examine the lowest energy annealed configurations and write down idealized variational spin states. The variational energy sits systematically below the MC energy because of the small-but-nonzero temperature. The phase diagram in the main text is determined from the level crossings over the entire $(\alpha,\lambda)$ plane, and a representative cut of this MC data at $\lambda=1$ with energy crossings is presented below. Our calculations were performed on a $30\times 30$ lattice, but there is a region of stability that supports a $4\times 1$ site commensurate Spiral-2 phase. Since this doesn't fit exactly on the lattice, the variational state has a domain wall at the lattice boundary. Thus, we somewhat underestimate the parameter region occupied by this phase. This could of course be corrected by utilizing a larger lattice for the simulations, but does not qualitatively affect our results.

The likely incommensurate spiral regions are not well described by any finite lattice method, but since the variational energies of the commensurate phases all have a roughly parabolic dependence on $\alpha$, we present this region of more smoothly varying energy in the main text as the possibly incommensurate Spiral-1 and 2 phases.

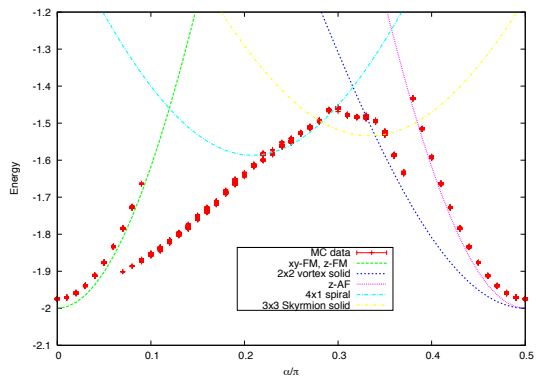

FIG. 1: (Color online) Unbiased MC annealed energies and variational energies of a variety of states at various $\alpha$ for $\lambda=1$. This illustrates the level crossings that ultimately result in the phase diagram in the main text.

**DETAILS OF THE INHOMOGENEOUS MEAN-FIELD THEORY FOR THE MAGNETIC SUPERFLUIDS**

A mean-field theory of the Mott transition begins with a decoupling of the kinetic part of the Hamiltonian by making the substitution $a_{i\sigma} = \phi_{i\sigma} + \tilde{a}_{i\sigma}$, where $\phi_{i\sigma} = \langle a_{i\sigma}\rangle$, and $\tilde{a}$ is the fluctuation around the equilibrium order parameter. For notational convenience, we also define $\phi_i^\dagger \equiv (\phi_{i\uparrow}^*, \phi_{i\downarrow}^*)$. Neglecting the terms quadratic in the fluctuations leaves a sum over single-site Hamiltonians:

$$H_{\mathrm{MF}} = \sum_i H_i - h_i^\dagger \cdot \psi_i - \psi_i^\dagger \cdot h_i + \phi_i^\dagger \cdot h_i \qquad (8)$$

where $H_i$ contains the exact local interaction and chemical potential terms, and the $h_i$ are classical spinor local fields that arise on each site through its coupling to neighbor sites,

$$h_i^\dagger = -t \sum_\mu \phi_{i+\mu}^\dagger \mathcal{R}_{i+\mu,i} \qquad (9)$$

on the 2d square lattice, $\mu$ takes on the values $\pm\hat{x}, \pm\hat{y}$. This single-site Hamiltonian $H_{\mathrm{MF}}$ can be diagonalized exactly, but is coupled through the order parameter fields to neighboring sites. A homogeneous approximation is inappropriate, so we use a finite cluster of sites ($8 \times 8$ in the present work) with periodic boundary conditions and perform a self-consistent, iterative minimization. Given a local environment, we find the lowest energy eigenstate on site $i$ and calculate $\langle a_{i\sigma}\rangle$ and store the value. Updating site $i$ in turn changes the value of $h$ for each of its neighbors, so we must perform several sweeps (on the order of $10^2$) over the entire cluster until we have convergence for all sites. To prevent settling into local minima, we repeat this process for several different random initial configurations.